\documentclass{article}

\usepackage{PRIMEarxiv}

\usepackage[utf8]{inputenc} 
\usepackage[T1]{fontenc}    
\usepackage{hyperref}       
\usepackage{url}            
\usepackage{booktabs}       
\usepackage{amsfonts}       
\usepackage{nicefrac}       
\usepackage{microtype}      
\usepackage{lipsum}
\usepackage{fancyhdr}       
\usepackage{graphicx}       
\usepackage{enumerate}
\usepackage{enumitem}
\graphicspath{{media/}}     
\usepackage{float}

\title{IoT Solution for Winter Survival of Indoor Plants}

\author{
  Md Saroar Jahan \\
  University of Oulu, CMVS, BP 4500, 90014 Finland\\
  \texttt{mjahan18@edu.oulu.fi} \\
   \And

  Jhuma kabir Mim\\
  LUT University, 53850 Lappeenranta, Finland\\
  \texttt{jhuma.mim@student.lut.fi} \\
  \And
  
  Sampo Niittyviita, Santeri Moberg, Murad Ahmad, Nijar Hossain \\
  University of Oulu, UBICOMP, 90014 Finland\\
  \texttt{\{saniitty, santeri.moberg, murad.ahmad, nhossain18 \}@student.oulu.fi} \\

}

\begin{document}
\maketitle

\begin{abstract}
Not only does cold climate pose a problem for outdoor plants during winter in the northern hemisphere, but for indoor plants as well: low sunlight, low humidity, and simultaneous cold breezes from windows and heat from radiators all cause problems for indoor plants. People often treat their indoor plants like mere decoration, which can often lead to health issues for the plant or even death of the plant, especially during winter. A plant monitoring system was developed to solve this problem, collecting information on plants’ indoor environmental conditions (light, humidity, and temperature) and providing this information in an accessible format for the user. Preliminary functional tests were conducted in similar settings where the system would be used. In addition, the concept was evaluated by interviewing an expert in the field of horticulture.
The evaluation results indicate that this kind of system could prove useful; however, the tests indicated that the system requires further development to achieve more practical value and wider usage.

\end{abstract}


\section{Introduction}

Several factors present challenges when growing plants at home in Finland. One of them being Finland’s climate with long and dark winters and short summers. Also, during spring and autumn, while the average temperature is typically above zero degrees during the day, the temperature can drop under zero degrees celsius during the night, which can inflict damage on plants outside. In addition, during the winter, there are only a few hours of daylight available which can be inadequate for some plants. The Nordic climate also presents challenges for indoor plants. Especially in the winter, some of the indoor plants must be placed next to the window where they can get enough light, yet that could also expose them to colder temperatures. A study by Nguyen et al. shows that absolute humidity correlates between outdoor and indoor air, which then makes the relative humidity drop in the indoors during cold weather \cite{nguyen2014relationship}. This drop-in relative humidity in winter should also be considered when growing plants indoors.

For greenhouses and indoor home environments, controlling the growing environment is crucial since the environment entirely relies on human control. The moisture of the soil, temperature, illuminance, and humidity are some of the environmental variables that a person can control for growing plants. However, determining the optimal growing conditions for certain plants can be tricky. This is especially the case for people who lack experience. Sometimes growing plants indoors is challenging because of the lack of proper knowledge about the plants the person is growing. To help small-scale indoor farming, including large-scale farming, the use of modern sensors and mobile applications could be a great addition.

In this research, a sensor system with an Android user interface was developed, which can be used to measure the environmental conditions of indoor plants. These sensors are able to collect temperature, humidity, and illuminance data surrounding the plant and show them in real-time to the user. Ìf the environmental conditions are outside the recommended range for a certain plant, the app will notify the user.

The use of a convenient mobile application could also help to promote small-scale agriculture by making it easier for new home growers to be successful in gardening. It could also improve the level of localization of food production.

\section{Related Work}
There are a number of different terminologies in use for smart agriculture (SA), including precision agriculture, variable rate technology, Global positioning system (GPS) agriculture, farming by inch, information-intensive agriculture, and site-specific crop management \cite{abbasi2014review,srinivasan2006precision}. The topic of smart agriculture has been researched since the 1990s \cite{srbinovska2015environmental}. It is a broad topic covering many different kinds of areas of research, from wireless sensor networks (WSN) to unmanned aerial systems (UAVs) \cite{zhang2012application}.

\subsection{Sensing in Smart Agriculture}
According to Ruiz-Garzia L. et al., the first wireless sensor network application for greenhouses was published in 2003 which included sensors for wind flow, wind direction, ambient light, temperature, ambient pressure, humidity and CO2 percentage \cite{ruiz2009review,liu2003application}. These are some of the variables that are commonly detected by sensors in SA solutions. Table 1 displays common variables that can be detected by sensors and are available in the market \cite{abbasi2014review,ojha2015wireless}. There are a wide range of market available sensors; many of which can collect data from  multiple environmental variables \cite{abbasi2014review}. Various systems have been developed to account for these variables. Examples of different applications are provided in the next chapter.

\begin{table}[H]
 \caption{Different environmental factors measurable by market available sensors.}
  \centering
  \begin{tabular}{|l|l|p{1.7cm}|p{1.7cm}|l|l|l|}
     \hline
    \multicolumn{7}{|c|}{\textbf{Soil}}                   \\
    \hline
    Temperature & Moisture & Dielectric  permittivity  & Rain/water flow & Water level & Conductivity & Salinity \\ 
    \hline
    \multicolumn{7}{|c|}{\textbf{Plant}}                   \\
    \hline
      Photosynthesis & Moisture & Hydrogen & Wetness & CO2 & Temperature & - \\ 
    \hline
    \multicolumn{7}{|c|}{\textbf{Weather}}                   \\
    \hline
    Temperature & Humidity & Atmospheric pressure & Wind speed   & Wind direction & Solar radiation  & Rainfall\\ 
     \hline
  \end{tabular}
  \label{tab:table}
\end{table}

\subsection{Smart Agriculture in Practice}
 A work by Khurshid Aliev and Eros Pasero presented a practical approach to acquiring data of temperature, humidity, and soil moisture of plants for smart agriculture system \cite{aliev2018internet}. They built a prototype device and an Android application that obtains physical information and sends it to the cloud to achieve the goal of their project. Moreover, they have focused on a temperature forecasting application in the subsequent part of their current research work.

Lashitha et al. presented an efficient sensor, and remote systems coordination of IoT in an actual agricultural circumstance \cite{priya2018smart}. By gathering ongoing information about agriculture that gives simple access to the farmer, their project's main objective. They developed task screening that utilizes different parameters so that it will help the farmer to screen more than one rural field in the meantime while doing their smart farming. Additionally, the vast majority of the observation is done remotely; it will help the person pick up data. As a result, observing through their framework requires less labor; individuals with physical handicaps can be utilized for checking fields.

G. Vellidis and M. Tucker have described the smart sensor array, and testing in a cotton crop \cite{vellidis2008real}. Inputs from the smart sensor array will determine timing and amounts for real-time site-speciﬁc irrigation applications while doing the integration of the sensors with precision irrigation technologies will provide a closed-loop irrigation system. That means the array consists of a centrally located receiver connected to a laptop computer and multiple sensor nodes installed in the ﬁeld, a specially designed circuit board and a Radio Frequency Identiﬁcation (RFID) tag which transmits data to the receiver for reliably monitoring spatially variable soil water status in crop ﬁelds.

Shailendra Wadje and Mahindra Gadakh proposed a Smart Farming (SF) System to increase the Production of plants \cite{wadje2017smart}. SF focuses on three main parts as Sensors, Microcontroller, and a system. They have focused on control, such as watering capturing the images based on statistical data sensed from sensor systems (including temperature, humidity, moisture, and light intensity sensors), by applying Kalman filtering to take more précised data before using an input into their decision-making process. A decision tree model has been used to predict data about the weather condition. After that, the set of decision-based data on both sensed data and weather conditions is developed automatically to decide on whether the watering and light system should be on or off. They also provided the function to manually control the watering and light system through a mobile application for smart farming.

\subsection{Issues in Smart Agriculture}
Smart agriculture is used for its potential toward improving agriculture. For example, automated agriculture sensing systems can be very effective compared to traditional methods; for instance, Gutiérrez J. et al. created a computerized system and tested it in a sage crop field for 136, which resulted in water saving of up to 90\% \cite{gutierrez2013automated}. Despite efficacy like this, there is yet room for improvement.

Review papers that focus on different aspects of SA state a wide range of issues related to the application of SA. Issues derive from the state of technology associated with SA, currently available systems in the market, and the farmers themselves. As stated in the paper by Mare Srbinovska et al., current issues include poor utilization of new technologies related to smart agriculture, which is caused by lack of investments and unfavorable decision-making \cite{srbinovska2015environmental}. According to Srbinovska et al., poor decision-making is due to a lack of information regarding the potential of using intelligent agricultural solutions for farming. Many issues also arise from the use of big data in agricultural systems, which S. Wolfert et al. state in their review paper of big data in smart farming:

\begin{enumerate} [topsep=2pt,itemsep=2pt,partopsep=2pt, parsep=2pt, label=\alph*)]
\item Data ownership and related privacy and security issues,
\item data quality,
\item intelligent processing and analytics,
\item sustainable integration of big data,
\item attractive business models which enable fair share between different stakeholders, and
\item openness of platforms \cite{wolfert2017big}.
\end{enumerate}

As current issues related to the state of the technology, low battery power, limited computation capability, and small memory of the sensor nodes are mentioned \cite{ojha2015wireless}.  Most of these issues also affect individuals outside the farming industry if they wished to start using similar solutions in their gardening or farming. Currently, the focus of SA applications are on large scale agriculture; thus, SA could be effectively applied to smaller scale as well. One area of improvement in the management of the plants in SA is to have an individualized approach for each plant to meet their needs in different environmental conditions \cite{mulla2013twenty}.

\begin{figure}
    \centering
    \includegraphics[width=.6\linewidth,]{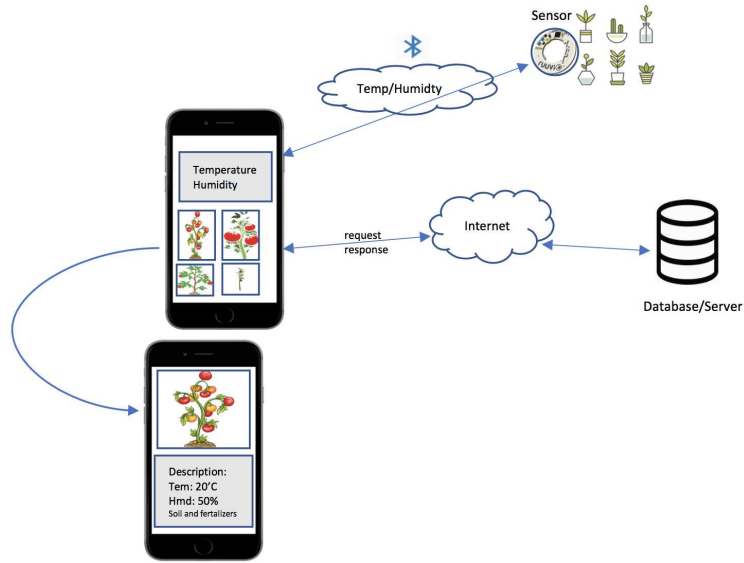}
    \caption{Application architecture.}
    \label{fig:1}
\end{figure}

\section{Design}
In this section, the design of the application and its parameters are presented. While this application can be beneficial even for an experienced gardener, the main target group for this application is the general public who are interested in growing plants as a hobby but do not have sufficient information about plants or their optimal living environmental conditions. The basic idea of the application is to help plants survive winter in indoor locations in good health and provide suggestions to improve the conditions of each plant species that the user might be growing. The system consists of a bluetooth sensor device called ruuvitag and an Android smartphone to which it is connected. The application is designed to get temperature and humidity values from the ruuvitag and illuminance data from the smartphone’s built-in light sensor, showing the values in real-time to the user. Additionally, the past sensor data can be viewed in the form of a line graph and a CSV file. Based on the sensor data, the application sends notifications to the user if their added plants are out of their optimal range in terms of temperature, humidity, or light. The software architecture for the application is presented in Figure~\ref{fig:1}.

\section{Implementation}
An Android application, which provides feedback on how the user could improve the living conditions of their plants, was developed for the Android platform. Since the user could have multiple plants in a single plant pot, the application has support for multiple plants. The purpose for that is to give specialized information for each of the plants the user might have. For example, some plants may cope with less sunlight during the winter; some plants are more adapted to winter conditions. A notification system was implemented to inform the user if there are actions that they should take to ensure a healthy environment for their plants. For each plant, there is set an optimal range for temperature, relative humidity, and illuminance. If the values that are sensed by the sensors go out of the optimal range for a more extended period, the user receives a warning notification. These ranges are different for each plant species because every species have their own optimal range of suitable growing conditions. The data for the optimal living conditions for a specific plant species is stored in a cloud database, where it can be fetched through the internet or exported in the desired format and manually transferred to the mobile device.

\subsection{Equipment and Software}

Android Studio was used for developing the application. Android mobile Motorola Moto G (3rd gen) and Google Pixel were used to test the software (Motorola Moto G (3rd gen) in evaluation). Java programming language was used for development (Android programming). The system was developed on Android version 8.0.0. The compile SDK version was 28, and the minimum SDK version was 18, along with some libraries including android-beacon, bluetooth, gson, and volley. Ruuvitag was used for sensing relative humidity and temperature. SQLite was used to store the sensor data and information regarding the plants locally on the mobile device.  

\subsection{Application}

Sensor data is transferred directly via bluetooth from the ruuvitag to an Android smartphone. Based on the data, the user gets real-time information regarding the precise environmental conditions of the plants in the sensor device’s close proximity. The user has added in the app for each individual plant species; he/she will get notified if the environmental conditions are not within their optimal range. Users can remove or add new plants to be monitored by the app. Based on the notifications the user gets from the application, the user can then do the corrections to the plants’ environment.

Data logging was implemented by writing the values of temperature, humidity, and illuminance along with a timestamp to a CSV file every few seconds. This data was later used in the evaluation of the application.

The application has the following components:

\begin{enumerate} [topsep=2pt,itemsep=2pt,partopsep=2pt, parsep=2pt, label=\alph*)]
 \item UI,
\item API for Ruuvitag, 
\item data logging,
\item plant management,
\item database and
\item notification system.
\end{enumerate}

\subsection{User Interface (UI)}

\begin{figure}[H]
    \centering
    \includegraphics[width=.6\linewidth,]{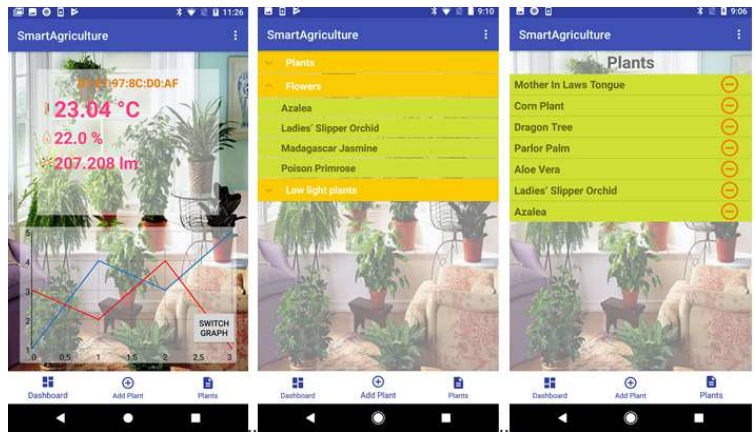}
    \caption{Screenshots of the smartphone UI.}
    \label{fig:2}
\end{figure}

There are three main screens of user interface as shows in Figure~\ref{fig:2}:
\begin{enumerate} [topsep=2pt,itemsep=2pt,partopsep=2pt, parsep=2pt, label=\alph*)]
    \item Main screen,
    \item Add plant, and
    \item Plant management
\end{enumerate}

On the main screen, the current environmental conditions are displayed, and bottom navigation is used for other screens. When the application is launched, the main screen displays the real-time temperature values, humidity, and illuminance. Below the sensor data on the main screen, a graph shows the history of the sensor values to the user. 

In the bottom navigation, there are three buttons to access each of the screens: "Dashboard" for the main screen, "Add plant" for adding the plants, and "Plants" for the plant management. In the "Add plant" screen, the plants are divided into categories, and different plants are shown, which are stored in the database. Users can add a plant from the list of plants to their existing plants by tapping on the plant's name. The user can see the existing plants that they have added to the application in the plant management screen. Users can also remove a specific plant from their existing plants by pressing the minus-icon.

\subsection{Plant Database}
The database was implemented to contain eight properties for each plant, which include:
\begin{enumerate} [topsep=2pt,itemsep=2pt,partopsep=2pt, parsep=2pt, label=\roman*)]
    \item Name of the plant,
\item Maximum temperature (Celsius),
\item Minimum temperature (Celsius),
\item Maximum humidity (relative humidity),
\item Minimum humidity (relative humidity),
\item Maximum illuminance (lux),
\item Minimum illuminance (lux), and
\item Description of the plant.
\end{enumerate}

The maximum and minimum values determine the optimal range for the variable in question regarding the plant's growing condition.

The plan was to gather data for each plant, searching the internet for guides and advice about growing the selected plant. Because of time constraints, data only for one plant species was gathered, Peace Lily (Spathiphyllum) plant, used in the evaluation part. Figuring the optimal growing conditions for Peace Lily was done by searching the internet for information on how to treat a Peace Lily plant. Multiple sites stated that the optimal temperature for Peace Lily is between 18 to 25 degrees Celsius. Since Peace Lily is a tropical plant, it also likes high humidity (40-90\%). If the humidity is under 40\%, it might cause health problems for the plant. Determining the proper lux values for the plant proved to be infeasible; thus, only textual  information was recorded for the amount of light the Peace Lily needs. Which was "low to moderate light". Data for the Peace Lily plant was stored locally on the installed app (hardcoded) using SQLite.

\section{Evaluation}

The purpose of the application is to develop a tool for helping plants survive winter in living apartments. The evaluation was performed to see how the application performs in this task; thus, the goal of the evaluation is to measure: how accurate and meaningful data is presented for the user. The assessment consisted of two parts: (i) concept evaluation and (ii) functional evaluation. The concept evaluation was performed in the form of an interview for a single professional in the field of horticulture to remediate potential issues that were not taken into account in the implementation and get some feedback on the aspects of the system. The functional evaluation was conducted to test the functionalities of the core system features and data reliability.

\subsection{Interview Setup}
The interview was conducted at the University of Oulu for a professional horticulturist with 33 years of experience with plants. The language used in the interview was Finnish since that was the native language of the interviewee. The whole interview was voice recorded and later analyzed and transcribed into English. The interview was conducted in two parts. The first part consisted of questions regarding the interviewee's background, and in the second part, the concept of the application and the work-in-progress application were presented to the interviewee. After that, the interviewee was asked to fill an AttrakDiff questionnaire. This was followed by questions regarding the concept and more general questions about plant care during winter.

\subsection{Functional Evaluation Setup}
Two peace lily (Spathiphyllum) plants which are common house plants in Finland, were used in two different locations to conduct the evaluation. This was done to give variability to the potential plant effects which might affect the measurements. Real plants were used because the plants and their pot might affect the measurements; for example, the plant might provide shade for the measurement device and affect the measurements' moisture levels or temperature. The plants were bought from a local nursery garden. Also, the well-being of the plant might give some indications regarding the plants' conditions. The plant pots' original material was plastic which was the material of the pots the plants were sold in. This plant pot was used in the evaluation.

\begin{figure}
    \centering
    \includegraphics[width=.7\linewidth,]{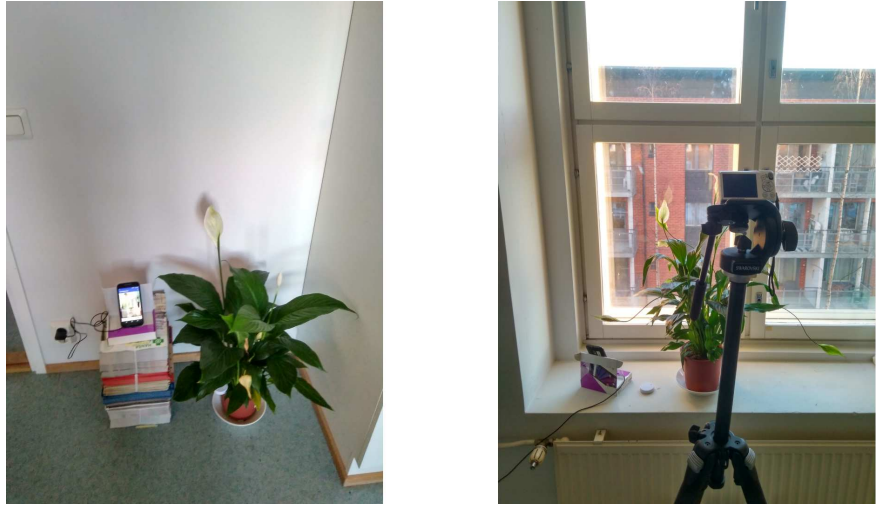}
    \caption{Two locations used in the experiment.}
    \label{fig:3}
\end{figure}

The experiment setup consisted of both plants, where the sensors were placed next to the plants. The plants were located in two different rooms in the same apartment. Figure 3 presents the experimental setup in the locations: the corner location and the window location. The windows of each room were directed towards different compass points, and the plants were located at a slightly different height from the floor. For sensing illuminance, the smartphone was placed at an angle of 45 degrees so that the sensor would receive a similar amount of light as the plant next to it would. The ruuvitag sensor device was placed under the leaves to mimic the humidity and temperature that the plants' experience (the plants change air under their leaves). The evaluation was conducted between 24.11.2018 and 8.12.2018 (14 days). For the whole period, the Android smartphone (Motorola Moto G 3rd Gen) was constantly charged and within the bluetooth range of the sensors. 

In the evaluation, the following actions were taken to gather data from the evaluated system:

\begin{itemize}
\item Displayed sensor data every few seconds (humidity, illuminance and temperature),
\item Pictures of the plants.
\end{itemize}
Thus, the focus of the evaluation was on the sensor data and not on the usability of the system. The pictures of the plants were taken every day from the same time of the day from the same angle and distance to see if the plants had any observable changes in their appearance during the experiment. The collected sensor data from the devices was uploaded every day to external storage in CSV format; moreover, the plants were observed carefully on a regular basis, and the living environment was kept the same.

Before the start of the experiment, the leaves of both of the plants were cleaned with a wet towel to remove any dust which might have an effect on their photosynthesis. Camera and sensor locations were marked with tape on the ground to ensure that they were not moved during the experiment. Plant care guidelines were followed to ensure plant health: fertilizers were not used since fertilizing during winter could potentially kill the plant, and each plant was given ten dl amount of water every week.

\section{Evaluation Results}
The concept of the application appeared to be promising based on the feedback gained from the interview. For functional evaluation, sensor data during the experiment was gathered. Data showed that the system can be used to measure temperature, humidity, and illuminance within the plants vicinity and that there is a meaningful difference in the sensor values between two different locations in the same apartment. Also, the sensor values being outside the optimal range for the test plants showed how this kind of application could be helpful.

\subsection{Interview Results}
The interview gave insight into how smart agriculture applications mostly focus on the greenhouse environment. There are not many, if any, applications that are for the hobbyist to use, which was also hypothesized based on the state-of-the-art research. Also, sensor systems used in a greenhouse do not fit well in a home environment because the spectrum of different kinds of house plants is much larger than in a greenhouse where there are many plants of the same species, to which the system is calibrated.

Based on the interview, the software concept appeared to be promising with its' intended features. As an improvement suggestion, the interviewee stated that the concept could be broadened to be used even in summer, yet was this to be implemented; changes should be made to enable support for summer conditions and are considerably different. If there were more than one plant in a room, more sensors would be helpful, especially more light sensors.

The application could be useful for both hobbyists and casual plant owners. Dedicated hobbyists may require precise plant information from the plants day and night, especially in winter, while casual plant owners often neglect their plants due to a lack of information. The interviewee often receives customers who tell that their plants have died during the winter, which is not usually the case for more dedicated hobbyists.

The interviewee showed interest in using an application of this kind: he/she has not encountered this kind of application before. According to his/her, the existing greenhouse applications are very complex and cannot be used in practice by the average person. In addition to the measured conditions, light, humidity, and temperature, information regarding plant watering could be helpful since watering requirements are not the same during winter and summer.

The AttrakDiff questionnaire result was deemed inaccurate - the interviewee expressed his/her positive outlook on the software, yet without any nuances that might be lacking. He/she rated the software with the highest scores in the questionnaire. The AttrakDiff would be more useful with a larger sample size and in a situation where the participant could perform the questionnaire anonymously. Since he/she was the only participant in the concept evaluation, the participant might feel more pressured or vital, which might affect the results.

\subsection{Functional Evaluation Results}
The results show that the system can be used to measure light level, humidity, and temperature in the plants' vicinity. The evaluation results were for this short period; however, the pictures are taken every day during the evaluation every day and plants showed no sign of observable changes in either of the plants. This would indicate that at least for the period of the evaluation, the plants' health did not deteriorate significantly, even though conditions were not optimal for a Peace Lily plant. The temperature in the window location was under the recommended 18-25 degrees celsius, and humidity was under the recommended 40-90\% humidity in both locations. The sensor data from both locations consisted of temperature, humidity, and illuminance values and a timestamp that was saved in a CSV file. This file was used for the analysis.

\begin{figure}
    \centering
    \includegraphics[width=.6\linewidth,]{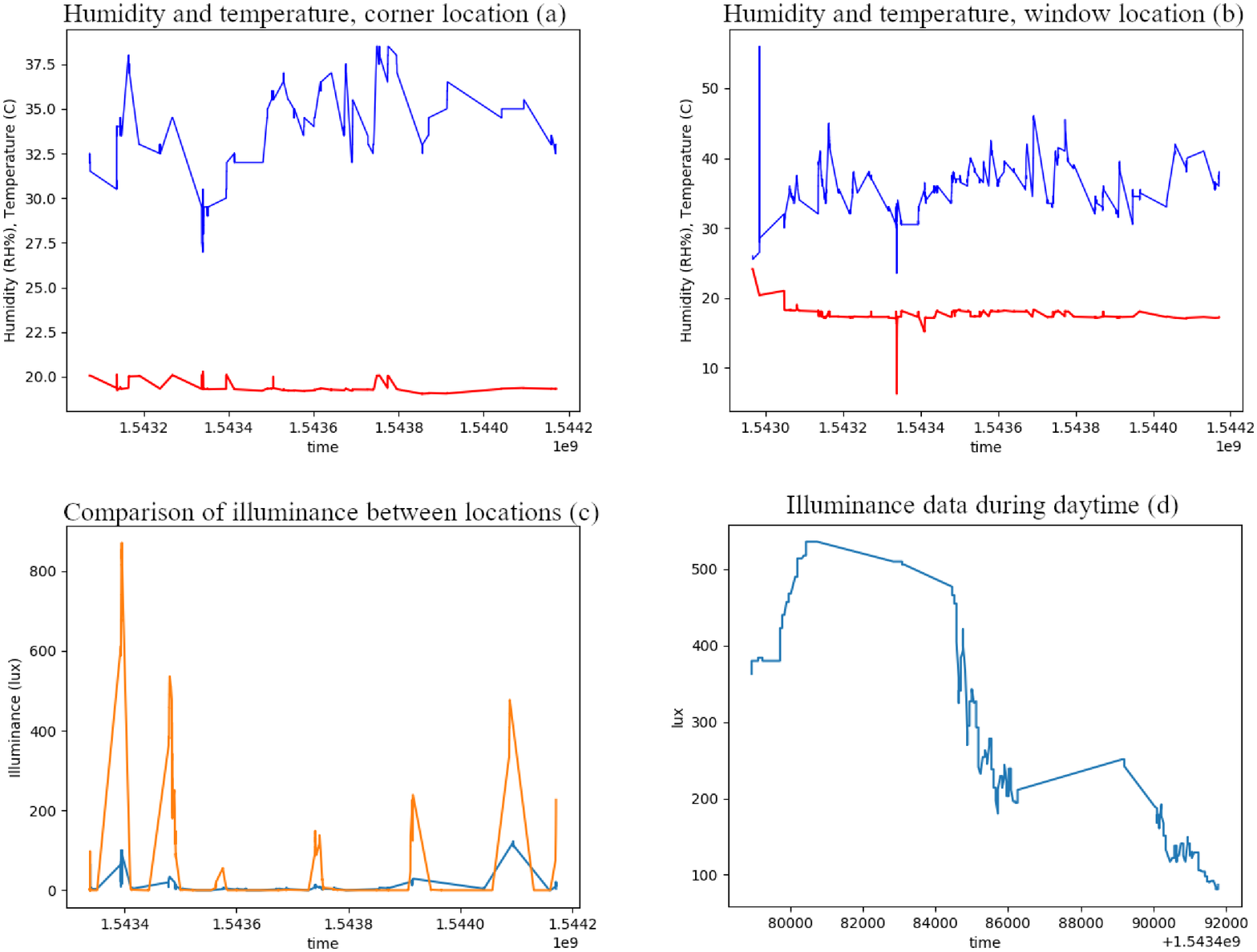}
    \caption{Line graphs from the collected sensor data.}
    \label{fig:4}
\end{figure}

Figure~\ref{fig:4} presents line graphs of temperature and humidity values during the experiment in both locations, comparing the illuminance in both locations and illuminance during daytime in one day. We can see from the collected data that the temperature was the steadiest of the three variables; there is only little variance, and only one spike of cold temperature can be seen during the experiment. That could have been due to a cold draft made by opening a window, for example. Relative humidity varied between 30 to 45 in both of the locations, which is outside of the recommended humidity for the Spathiphyllum plant. The illuminance was near zero during most of the experiment and only spiked during the daytime. The highest value of illuminance recorded was 871 luxes. During this peak, the illuminance value in the location further away from the window was only 70 luxes, which shows that distance from the window affects a lot on how much light the plant gets.

Average sensor values during the experiment in both of the locations are presented in Table 2. The temperature was almost two degrees warmer in the window corner location, which was further away from the window. The difference is significant because the temperature in the window location lower than the recommended temperature for the Peace Lily plants that were used in the experiment. The difference in location does not seem to affect much in terms of humidity, the only difference of 1,62 percentage points. Humidity was also lower than the recommended humidity for Peace Lily plants in both of the locations.  Illuminance had the biggest difference between the locations. The average illuminance in the window locations was over seven times as much as in the corner location.

\begin{table}
 \caption{Average sensor values during experiment.}
  \centering

\begin{tabular}{ |p{2cm}| p{2cm} |p{2cm} |p{2cm}| } 

\hline
 Location & Temperature & Humidity & Illuminance \\ 
 \hline
 Corner & 19.48 °C & 34.24  & 10.36 lux \\ \hline
 Window & 17.59 °C & 35.86   & 75.55 lux\\ 
\hline
\end{tabular}
\label{tab:table_2}
\end{table}

\subsection{Limitations}
During the two-week experiment, the application that was running on both of the smartphones crashed multiple times. While the application was down, no sensor data was gathered, and the application had to be restarted manually. This resulted in significant gaps in the collected sensor data. Also, it is to be noted that the evaluation only had two plants in two locations - it would be good to have more plants and more locations to get more reliable and consistent. Due to these points, no generalization should be made from the gathered data.

The tested system required a constant connection between the Android smartphone and the sensor; however, this proved to be an insufficient solution; instead, the system should be implemented using either an external server node which would send data to a cloud database or the system would store the data from longer periods until it can be synced to the Android smartphone.

There are advantages and disadvantages for both multiple and single sensor approaches. A single sensor approach for all variables would be easier to manage yet slightly more inaccurate due to un-optimal placing (e.g., optimally, the illuminance is measured from the top of the plant and humidity under the plant's leaves). The system should be tested with both approaches on actual end-users to determine each approach's applicability. While, in theory, multiple sensors would give more accurate data - too many sensors might be too cumbersome to handle for normal users.

Since the experimenters did not have a background in gardening, the mishandling of the plants was an issue for consideration. The initial condition of the plants was that of the plant's condition, which they were available from the plant nursery. While it can be expected that the plants are well taken care of in the plant nursery, this might not always be the case, yet the plants appeared to be in good condition upon purchase. The condition of the plants is displayed in Figure 3. Also, in an actual situation, the mobile device may not necessarily constantly be within the range of the sensors; this would be relevant to include in the case of an available product.

\section{Future Works}
Another option for the implementation would be to include a network node that would have access to the internet. This would allow remote and constant access to the plants’ conditions; however, it would make the system more expensive. The number of devices required for this would increase from two to three, which would make the system more complicated to use and upkeep.

Instead of only one set of sensors, multiple sensors of the same type could be used. Yet, there are advantages and disadvantages for multiple and single sensor approaches. A single sensor approach for all variables would be easier to manage, yet potentially slightly more inaccurate due to un-optimal placing in relation to some of the plants. It would be useful to test the system with both approaches on actual end-users to determine the applicability of each of the approaches. The results also suggest that the use of multiple sensors would be beneficial if there are plants in multiple locations in the same apartment, as the measured locations had clearly different conditions.

As there are a large number of different plant species, the plant database would have to be very large to be all-encompassing. Optimally, for saving plant data and analyzing user data more effectively, a cloud server would be a good solution. Since gathering plant data would be very difficult to achieve, another solution could be investigated. One option for this would be to implement a parallel crowdsourced database, where the users would themselves add relevant plant information to the database. Additionally, other environmental variables which might affect the measurement include spatial requirements of the plants and watering of the plants and soil conditions. Of these measurements, the watering of the plants came up in the interview. While data from these variables would be interesting, these are not the most relevant variables for winter survival. It would make the system even more complex, less approachable, and expensive for a layperson. 

The messaging system in the implementation included two thresholds for messages; a low and a high, as discussed in plant database section 4.4. For each measured variable, it would be beneficial to have more thresholds instead of only low and high, for example, very low and very high in addition.

\section{Conclusion}
The results indicate that the system 1) can be used to measure light level, humidity, and temperature in the plant's vicinity, and 2) the measurement of these variables is useful. These preliminary results promise to conduct further research on the system and the area related to the system. The developed system would require more iterations. The information for optimal growing conditions would have to be accurate for a larger number of plants before the system would be sufficient for commercial use. 

The results also suggest that using multiple sensors would benefit plants in multiple locations in the same apartment, as the measured locations had clearly different conditions. The interview results also suggest this improvement. In addition, the concept could be broadened and some changes to be implemented, for example, for summer usage as the interview results suggested, yet, were this to be implemented, changes should be made to enable support for summer conditions as they would significantly differ from winter conditions.

\section*{Acknowledgments}
We would like to thank University of Oulu Ubicom Department for providing research material.

\bibliographystyle{unsrt}  
\bibliography{references}

\end{document}